\documentclass[conference]{IEEEtran}

\usepackage{graphicx}
\usepackage{color}
\usepackage{placeins}
\usepackage{float}
\usepackage{tabularx,colortbl}
\usepackage{amsmath}
\usepackage{multirow}
\usepackage{hyperref}
\usepackage{mathtools}
\usepackage{algorithm}
\usepackage{algorithmic}

\begin{document}
%
\title{Self-Interference Cancellation with Nonlinear Distortion Suppression for Full-Duplex Systems}


\author{\IEEEauthorblockN{Elsayed~Ahmed and Ahmed~M.~Eltawil\\}
\IEEEauthorblockA{Electrical Engineering and Computer Science\\
University of California, Irvine, CA, USA\\
\{ahmede,aeltawil\}@uci.edu \\}
\and
\IEEEauthorblockN{Ashutosh~Sabharwal\\}
\IEEEauthorblockA{Electrical and Computer Engineering\\
Rice University, Houston, TX, USA\\
ashu@rice.edu}}


\maketitle

\begin{abstract}
In full-duplex systems, due to the strong self-interference signal, system nonlinearities become a significant limiting factor that bounds the possible cancellable self-interference power. In this paper, a self-interference cancellation scheme for full-duplex orthogonal frequency division multiplexing systems is proposed. The proposed scheme increases the amount of cancellable self-interference power by suppressing the distortion caused by the transmitter and receiver nonlinearities. An iterative technique is used to jointly estimate the self-interference channel and the nonlinearity coefficients required to suppress the distortion signal. The performance is numerically investigated showing that the proposed scheme achieves a performance that is less than 0.5dB off the performance of a linear full-duplex system.
\end{abstract}


%
\IEEEpeerreviewmaketitle

\section{Introduction}
Full-duplex transmission is the communication scheme where bidirectional communications is carried out over the same temporal and spectral resources~\cite{Ref1}-\cite{Ref11}. The main limitation impacting full-duplex transmission is managing the strong self-interference signal imposed by the transmit antenna on the receive antenna within the same transceiver. Throughout the literature, several combinations of passive and active self-interference cancellation schemes have been proposed~\cite{Ref1}-\cite{Ref7}; aiming to mitigate the self-interference signal below the noise level. However, the experimental results in~\cite{Ref1}-\cite{Ref5} have demonstrated that complete self-interference elimination is not possible in current full-duplex systems, mainly due to a combination of system imperfections, especially radio circuits' impairments.

In order to understand the system limitations, several recent publications~\cite{Ref10}-\cite{Ref13r} have considered the problem of full-duplex transmission to investigate the impact of radio circuit impairments on the system performance and explore system bottleneck. More specifically, the results in~\cite{Ref13} show that, due to the large power differential between the self-interference signal and the signal-of-interest, system nonlinearity becomes one of the main factors that limit self-interference mitigation capability. Generally, system nonlinearity introduces in-band nonlinear distortion to the transmitted and received signals. Most of the existing self-interference cancellation schemes ignore the nonlinearity effect, which limits the amount of cancellable self-interference power to the distortion level.

In this paper, we consider the problem of self-interference cancellation in full-duplex orthogonal frequency division multiplexing (OFDM) systems in the presence of four main radio impairments: (i) transmitter and receiver nonlinearity, (ii) transmitter and receiver oscillator phase noise, (iii) receiver Gaussian noise, and (iv) analog-to-digital converter (ADC) quantization noise. A digital-domain self-interference cancellation scheme that accounts for the transmitter and receiver nonlinearity effect is proposed. The proposed scheme increases the amount of cancellable self-interference power by suppressing the nonlinear distortion associated with the received self-interference signal.

Suppressing the nonlinear distortion requires the self-interference channel as well as nonlinearity coefficients to be estimated. However, due to the presence of the nonlinear distortion while the self-interference channel is being estimated, the channel estimation error will be distortion limited. To overcome this problem, we propose an iterative technique to jointly estimate the self-interference channel and the nonlinearity coefficients required to perform self-interference cancellation and distortion suppression. The performance of the proposed scheme is numerically investigated and compared against the case of a linear full-duplex system. The results show that after three to four iterations, the nonlinear distortion is significantly suppressed such that the proposed scheme achieves a performance that is less than 0.5dB off the performance of a linear full-duplex system.

The remainder of the paper is organized as follows. In Section II, the signal model is presented. The proposed scheme is introduced in Section III. Simulation results and discussions are presented in Section IV. Finally, Section V presents the conclusion.

\section{Signal Model}
\begin{figure}[t]
\begin{center}
\noindent
  \includegraphics[width=3.5in,trim= 0in 0in 0in 0in]{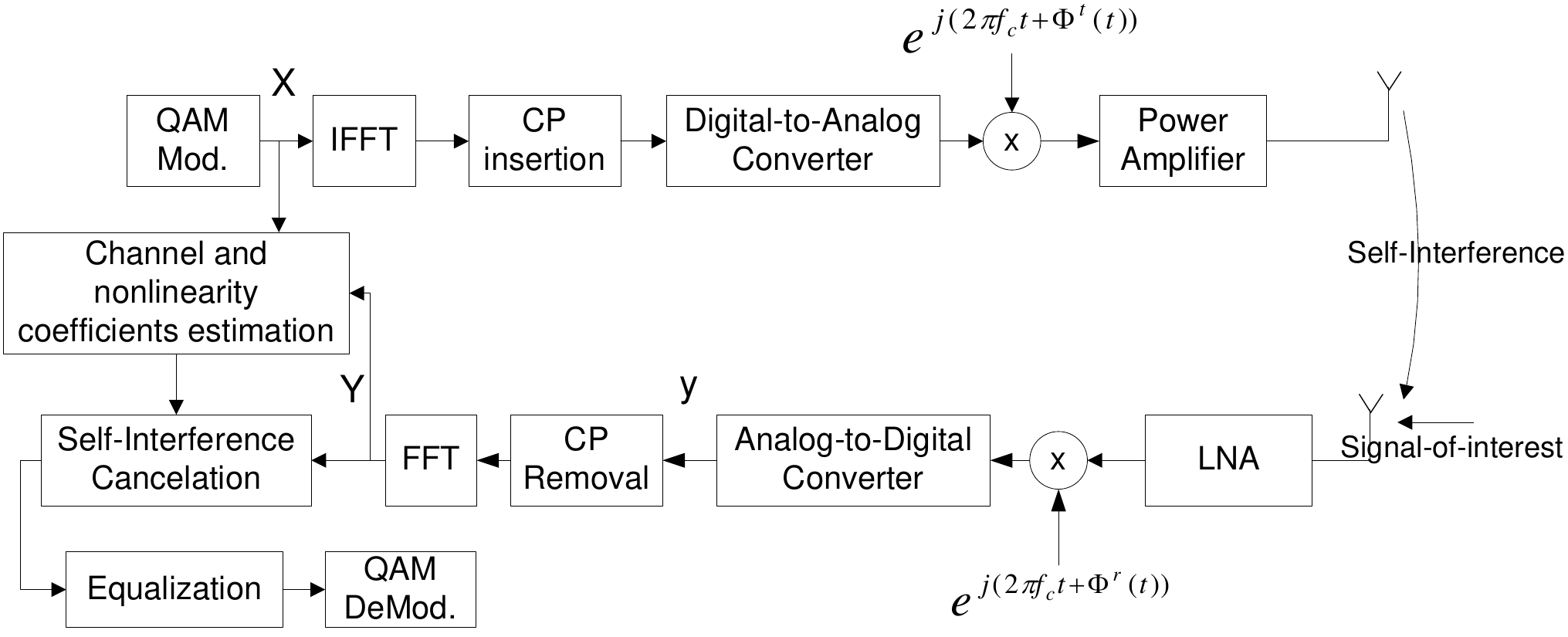}
  \caption{Block diagram of a full-duplex OFDM transceiver.\label{Fig1Label}}
\end{center}
\end{figure}

Figure~\ref{Fig1Label} illustrates a block diagram for a full-duplex OFDM transceiver, where the transmitter and the receiver are operating simultaneously over the same carrier frequency. At the transmitter side, the base-band signal is modulated using an OFDM modulator and then up-converted to the carrier frequency $f_c$, then amplified using a power amplifier. The oscillator at the transmitter side is assumed to have a random phase error represented by $\phi^t(t)$. At the receiver side, the amplitude of the received signal is properly adjusted using a low-noise amplifier (LNA). The signal is then down-converted from the carrier frequency to the base-band. The down-conversion mixer is assumed to have a random phase error represented by $\phi^r(t)$. The base-band signal is then quantized and converted to the frequency domain using Fourier transform.

In practical systems, the main sources of the system nonlinearity are the power amplifier at the transmitter side and the LNA at the receiver side. In this paper, we consider both the power amplifier and LNA nonlinearities. Generally, for any nonlinear block, the output signal $y$ can be written as a polynomial function of the input signal $x$ as follows~\cite{Ref14} 
\begin{equation}\label{eq:1}
y = \sum_{m=0}^{M-1} \alpha_{m+1} x^{m+1} \text{.}
\end{equation}
It can be shown that for practical wireless systems~\cite{Ref14}, only the odd orders of the polynomial contribute to the in-band distortion. Furthermore, only a limited number of orders contribute to the distortion and higher orders could be neglected. In practical systems, the nonlinearity is typically characterized by the third-order intercept point (IP3), which is defined as the point at which the power of the third harmonic is equal to the power of the first harmonic~\cite{Ref15}. Accordingly, in this paper we limit our analysis to the third-order nonlinearity where the output of any nonlinear block can be simplified as 
\begin{equation}\label{eq:2}
y = x+\alpha_3 x^3 \text{,}
\end{equation}
assuming a unity linear gain (i.e. $\alpha_1 = 1$).

Following the block diagram in Figure~\ref{Fig1Label} and using the assumption that $e^{j\phi}=1+j\phi$, $\phi \ll 1$, the base-band representation of the received signal at the ADC output can be written as
\begin{equation}\label{eq:3}
y_n = x_n^I*h_n^I + x_n^S*h_n^S + d_n + \phi_n + q_n + z_n \text{,}
\end{equation}
where '$*$' denotes convolution process, $n$ is the sample index, $x^I$, $x^S$ are the transmitted self-interference and signal-of-interest respectively, $h^I$, $h^S$ are the self-interference and signal-of-interest channels, $d_n$ is the total transmitter and receiver nonlinear distortion, $\phi_n$ is the total phase noise, $q_n$ is the ADC quantization noise, and $z_n$ is the receiver Gaussian noise. The receiver Gaussian noise represents the noise inherent in the receiver circuits, and usually specified by the circuit noise figure, which is implicitly a function of the LNA gain~\cite{Ref15}.

Using the nonlinearity model in~\eqref{eq:2}, and ignoring the nonlinearity associated with the signal of interest because of its small power compared to the self-interference signal, the total distortion $d_n$ can be written as
\begin{equation}\label{eq:34}
d_n = \underbrace{\alpha_3^t \left(x_n^I \right)^3 * h_n^I}_{\text{Transmitter nonlinearity}} + \underbrace{\alpha_3^r\left(x_n^I*h_n^I + \alpha_3^t \left(x_n^I \right)^3 * h_n^I \right)^3}_{\text{Receiver nonlinearity}} \text{,}
\end{equation}
where $\alpha_3^t$, $\alpha_3^r$ are the transmitter and receiver third-order nonlinearity coefficients. Expanding~\eqref{eq:34} we get
\begin{eqnarray}\label{eq:4}
d_n &=& \alpha_3^t \left(x_n^I \right)^3 * h_n^I + \alpha_3^r \left(x_n^I*h_n^I \right)^3  \nonumber \\
& & + 3 \alpha_3^t \alpha_3^r \left(x_n^I*h_n^I \right)^2 \left(\left(x_n^I \right)^3*h_n^I \right) \nonumber \\
& & + 3 \alpha_3^r \left(x_n^I*h_n^I \right) \left(\alpha_3^t \left(x_n^I \right)^3*h_n^I \right)^2 \nonumber \\
& & + \left(\alpha_3^t \left(x_n^I \right)^3*h_n^I \right)^3 \text{,}
\end{eqnarray}

According to~\eqref{eq:34}, the main difference between the transmitter and receiver nonlinearity is that the transmitter nonlinearity affects the signal only while the receiver nonlinearity affects both the signal and the wireless channel. Also it has to be noted that, although only 3$^{rd}$ order harmonics are considered at both transmitter and receiver sides, the coexistence of the transmitter and receiver nonlinearity introduces 5$^{th}$, 7$^{th}$, and 9$^{th}$ order harmonics (the 3$^{rd}$, 4$^{th}$, and 5$^{th}$ terms in~\eqref{eq:4}). The 7$^{th}$ and 9$^{th}$ order harmonics are much smaller than other harmonics, thus can be ignored. Accordingly, the distortion signal can be simplified as
\begin{eqnarray}\label{eq:5}
d_n &=& \alpha_3^t \left(x_n^I \right)^3 * h_n^I + \alpha_3^r \left(x_n^I*h_n^I \right)^3  \nonumber \\
& & + 3 \alpha_3^t \alpha_3^r \left(x_n^I*h_n^I \right)^2 \left(\left(x_n^I \right)^3*h_n^I \right) \text{.}
\end{eqnarray}
Finally, the received frequency-domain signal can be written as
\begin{equation}\label{eq:6}
Y_k = X_k^I H_k^I + X_k^S H_k^S + D_k + \Phi_k + Q_k + Z_k \text{,}
\end{equation}
where $k$ is the subcarrier index, and upper-case notation refers to the discrete Fourier transform (DFT) of the corresponding time-domain signals.

\begin{figure}[t]
\begin{center}
\noindent
  \includegraphics[width=3.3in,trim= 0in 0in 0in 0in]{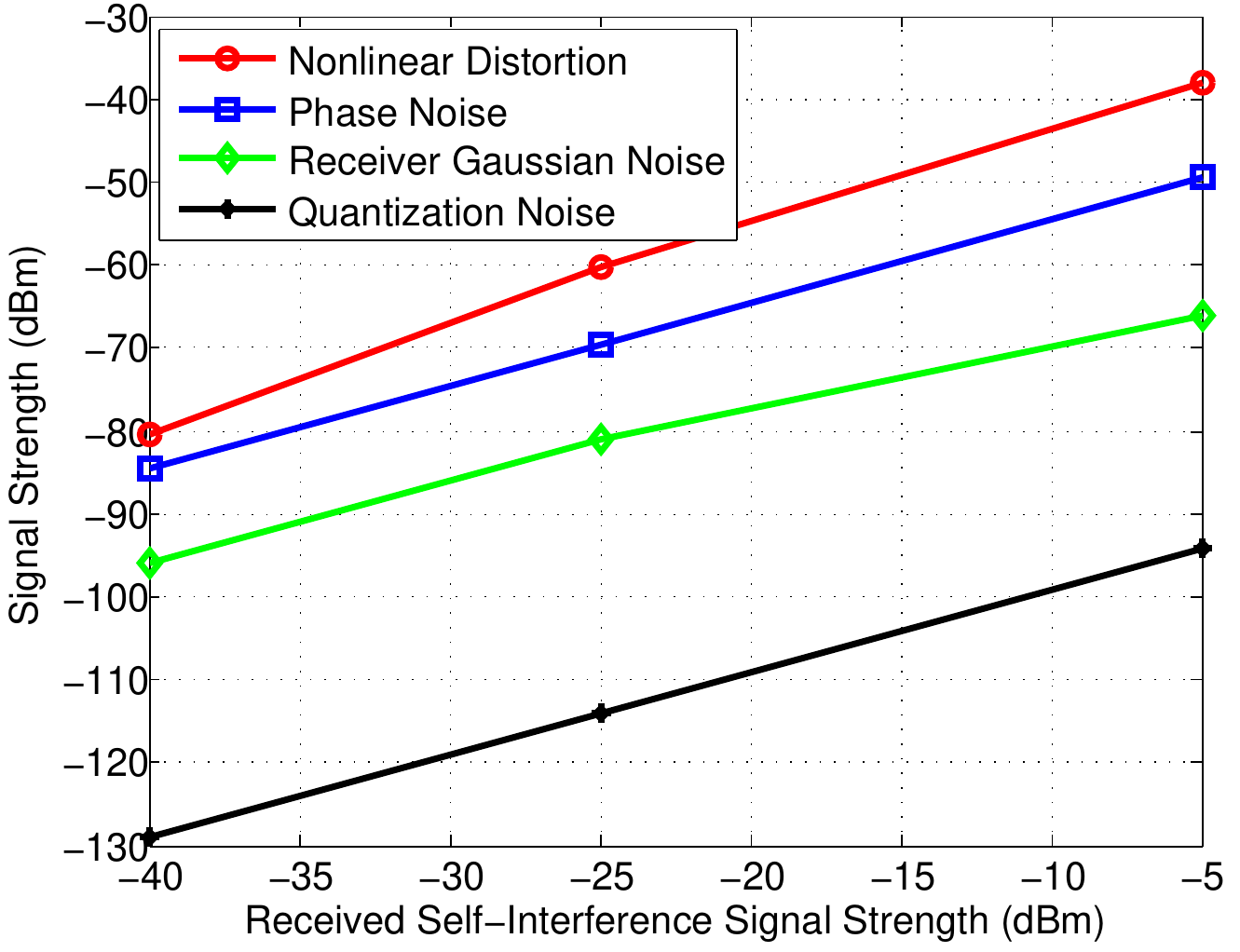}
  \caption{Noise powers at different received self-interference signal strengths for the transceiver in~\cite{Ref16}.\label{Fig2Label}}
\end{center}
\end{figure}

In order to show the significance of each noise term, the system is simulated using parameter values for a practical wireless transceiver~\cite{Ref16}. Figure~\ref{Fig2Label} shows the strength of each noise source at different received self-interference signal strengths. The results show that the nonlinear distortion is the main limiting factor, followed by the phase noise then the receiver Gaussian noise and quantization noise.

\section{Self-interference cancellation with distortion suppression}
The results in Figure~\ref{Fig2Label} imply that, eliminating the nonlinear distortion increases the self-interference mitigation capability. According to~\eqref{eq:5}, distortion elimination requires the knowledge of the self-interference channel ($h^I$) as well as the nonlinearity coefficients ($\alpha_3^t$, $\alpha_3^r$). In the proposed scheme, the self-interference channel is estimated using an orthogonal training sequence at the beginning of each transmission frame. The estimated channel along with the knowledge of the self-interference signal ($x^I$) are then used to estimate the nonlinearity coefficients.

\begin{figure}[t]
\begin{center}
\noindent
  \includegraphics[width=3in,trim= 0in 0in 0in 0in]{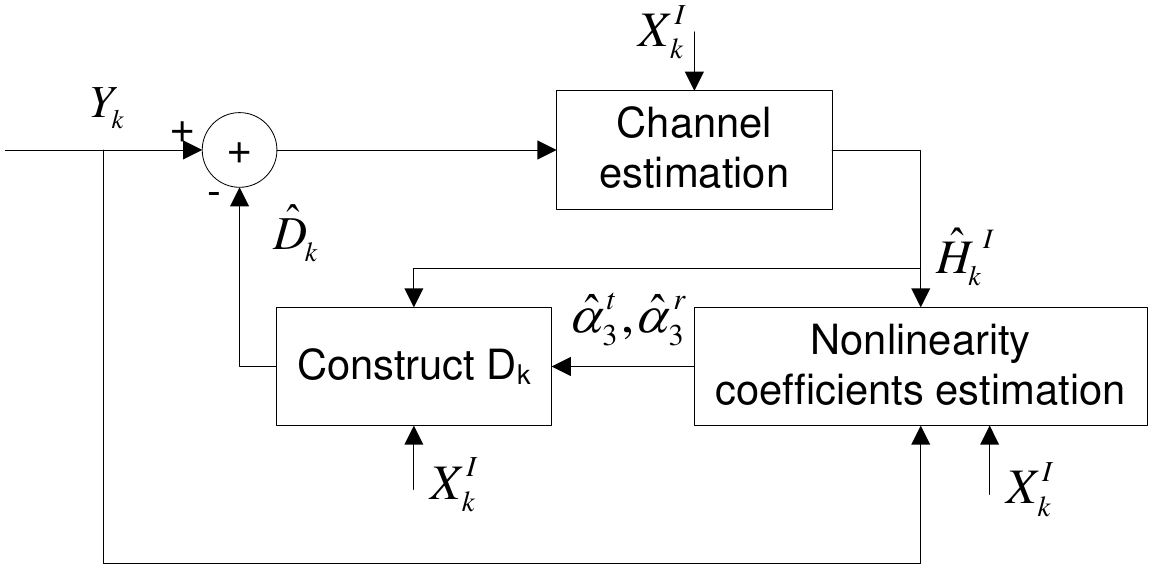}
  \caption{Block diagram for the iterative channel and nonlinearity coefficients estimation technique.\label{Fig3Label}}
\end{center}
\end{figure}

The main problem is that due to the presence of the distortion signal at the training time, the channel estimation error will be limited by the distortion signal, which impacts the estimation accuracy and thus the overall cancellation performance. To overcome this problem, we propose an iterative technique to jointly estimate the self-interference channel and the nonlinearity coefficients. The proposed technique consists of four main steps: (i) an initial estimate for the self-interference channel ($\hat{H}_k^I$) is obtained, (ii) the estimated channel is used to estimate the nonlinearity coefficients ($\alpha_3^t$, $\alpha_3^r$), (iii) the estimated coefficients are used to construct an estimate for the distortion signal $\hat{D}_k$, and (iv) the estimated distortion signal $\hat{D}_k$ is subtracted from the received signal. The four steps are then repeated for a number of iterations. An illustrative block diagram for the proposed iterative technique is shown in Figure~\ref{Fig3Label}.

After channel and nonlinearity coefficients estimation, the self interference signal ($X_k^I \hat{H}_k^I$) and the distortion signal ($\hat{D}_k^I$) are subtracted from the received signal at each data OFDM symbol to construct the interference-free signal. In the following subsections, detailed analysis for the channel and nonlinearity coefficients estimation techniques is presented.
\subsection{Channel estimation}
It has to be noted that for the iterative technique in Figure~\ref{Fig3Label} to work properly, the mean square error of the channel estimation should be less than the distortion power, otherwise the performance will be limited by the channel estimation error and there will be no gain achieved by the iterative technique. The DFT based channel estimation technique proposed in~\cite{Ref17} is one of the low complexity channel estimation techniques that achieve relatively small  mean square error. In this technique, first, an estimate for the channel impulse response (CIR) is obtained using the least square (LS) estimator as follows
\begin{equation}\label{eq:7}
\hat{h}_n^{LS} = \mathsf{IDFT} \left\{\frac{Y_k}{X_k} \right\} \text{.}
\end{equation}
Then, by leveraging the fact that the channel information is contained in the first $L$ samples of the CIR, a better estimate for the channel is obtained by taking the first $L$ samples of $\hat{h}_n^{LS}$ while forcing other samples to zero as follows
\begin{equation}\label{eq:8}
\hat{h}_n = \left\{
\begin{array}{c l}
    \hat{h}_n^{LS} & \text{,\ \ } 0 \leq n \leq L-1 \text{,}\\
    0 & \text{,\ \ otherwise} \text{,}
\end{array}\right.
\end{equation}
then
\begin{equation}\label{eq:9}
\hat{H}_k = \mathsf{DFT} \left\{\hat{h}_n \right\} \text{.}
\end{equation}
By doing this, the estimation error is reduced by a factor of $\frac{L}{N}$, where $N$ is the number of subcarriers per OFDM symbol. The key challenge in such technique is the choice of $L$. Since the cyclic prefix in practical systems is designed to be larger than the channel length, a good choice for $L$ is to be equal to the cyclic prefix length.
\subsection{Nonlinearity coefficients estimation}
At the self-interference training symbol, the signal-of-interest is not present. Therefore, Equation~\eqref{eq:3} can be written as
\begin{equation}\label{eq:10}
y_n = x_n^I*h_n^I + d_n + \phi_n + q_n + z_n \text{.}
\end{equation}
Since the transmitted self-interference signal $x_n^I$ and the self-interference channel $\hat{h}_n$ are now known, the problem in~\eqref{eq:10} can be recognized as a linear estimation problem with the unknown coefficients $[\alpha_3^t, \alpha_3^r, 3\alpha_3^t\alpha_3^r]$.

Rewriting~\eqref{eq:10} in a matrix form we get
\begin{equation}\label{eq:11}
\left[ \begin{array}{c} \bar{y}_0 \\ \bar{y}_1 \\:\\ \bar{y}_N \end{array} \right] =
\underbrace{\begin{bmatrix} A_1 & B_1 & C_1 \\ A_2 & B_2 & C_2\\:&:&: \\ A_N & B_N & C_N \end{bmatrix}}_{W}
\left[ \begin{array}{c} \alpha_3^t \\ \alpha_3^r \\ 3\alpha_3^t\alpha_3^r \end{array} \right] 
+ \left[ \begin{array}{c} \eta_0 \\ \eta_1 \\:\\ \eta_N \end{array} \right] \text{,} 
\end{equation}
where $\bar{y}_n=y_n-x_n^I*\hat{h}_n^I$, $\eta_n = \phi_n+q_n+z_n$, $A_n = (x_n^I)^3*\hat{h}_n^I$, $B_n = (x_n^I*\hat{h}_n^I)^3$, and $C_n = (x_n^I*\hat{h}_n^I)^2((x_n^I)^3*\hat{h}_n^I)$. Rewrite~\eqref{eq:11} in a compact form we get
\begin{equation}\label{eq:12}
\bar{y} = W \alpha + \eta \text{.}
\end{equation}
An estimate for the nonlinearity coefficients $\alpha$ can be found using the LS estimator as
\begin{equation}\label{eq:13}
\hat{\alpha} = W^{-1} \bar{y} \text{.}
\end{equation}

The main problem with the LS estimator is that the matrix $W$ is often ill-conditioned, thus the inversion of the matrix will incur numerical errors. To overcome this problem, we propose a successive one-by-one estimation technique to avoid matrix inversion. The proposed technique is similar to the successive interference cancellation technique where one coefficient (e.g $\alpha_3^t$) is estimated assuming that other two are equal to zero. The estimated coefficient is multiplied by its corresponding signal and subtracted from the received signal, then the next coefficient is estimated. Since the third coefficient ($3\alpha_3^t\alpha_3^r$) is function of the first two, estimating $\alpha_3^t$, and $\alpha_3^r$ is sufficient to get the three coefficients. Furthermore, for better estimation accuracy iterative techniques could be used.

A common problem with any successive technique is the determination of the coefficient to start with. If there is prior knowledge about the relative strength of the transmitter and receiver nonlinearity, the optimum choice is to start with the coefficient that corresponds to the stronger nonlinearity. For example if the transmitter nonlinearity is stronger than receiver nonlinearity, the algorithm should start with $\alpha_3^t$ and vise versa. However, if there is no prior knowledge, a wrong starting point might result in performance degradation. In order to overcome this problem, the proposed algorithm selects the start coefficient based on the residual distortion power. In other words, the coefficient that results in smaller residual distortion power will be selected as the start coefficient. The iterative successive nonlinearity coefficients estimation technique is summarized in algorithm~\ref{Alg1}. The equations in algorithm~\ref{Alg1} assumes that $\alpha_3^t$ is selected as the start coefficient. Finally, it has to be mentioned that, to compute $A_n$, $B_n$, and $C_n$ up-sampling is required in order to prevent aliasing 
\begin{algorithm}
\caption{Successive nonlinearity coefficients estimation}
\label{Alg1}
{
\begin{algorithmic}[1]
\STATE set $\bar{y}_n = y_n - x_n^I*\hat{h}_n^I$.
\STATE Determine the start coefficient based on the residual distortion power
\FOR{certain number of iterations}
\STATE get $\hat{\alpha}_3^t=\frac{1}{N} \sum_{n=0}^{N-1}\frac{\bar{y}_n}{A_n}$.
\STATE set $\bar{y}_n = y_n - x_n^I*\hat{h}_n^I - \hat{\alpha}_3^t A_n$.
\STATE get $\hat{\alpha}_3^r=\frac{1}{N} \sum_{n=0}^{N-1}\frac{\bar{y}_n}{B_n}$.
\STATE set $\bar{y}_n = y_n - x_n^I*\hat{h}_n^I - \hat{\alpha}_3^r B_n - 3\hat{\alpha}_3^r \hat{\alpha}_3^t C_n$.
\ENDFOR
\end{algorithmic}
}
\end{algorithm}

\section{Simulation results and discussions}
In this section, the performance of the proposed cancellation scheme is numerically investigated under different operating conditions. The simulation setup is chosen as in WiFi 802.11n standard~\cite{Ref18}. The indoor TGn channel model~\cite{Ref19} is used to model the self-interference and signal-of-interest channels. The self-interference and signal-of-interest channel's Rician factors are set to 30dB and 3dB respectively. Two performance criteria are chosen: the achievable rate, and the residual interference plus distortion plus noise (RIDN) power. The RIDN is calculated as
\begin{equation}\label{eq:14}
\mathsf{RIDN} = X_k^I \left(H_k^I - \hat{H}_k^I \right) + \left(D_k - \hat{D}_k \right)+ \Phi_k+Q_k+Z_k \text{.}
\end{equation}
The proposed algorithm is compared to two cases; first, the case of linear full-duplex system (the best case) where $D_k=0$. Second, the case of nonlinear full-duplex system and no distortion removal is performed (as assumed in most current cancellation schemes).

\begin{figure}[t]
\begin{center}
\noindent
  \includegraphics[width=3.3in,trim= 0in 0in 0in 0in]{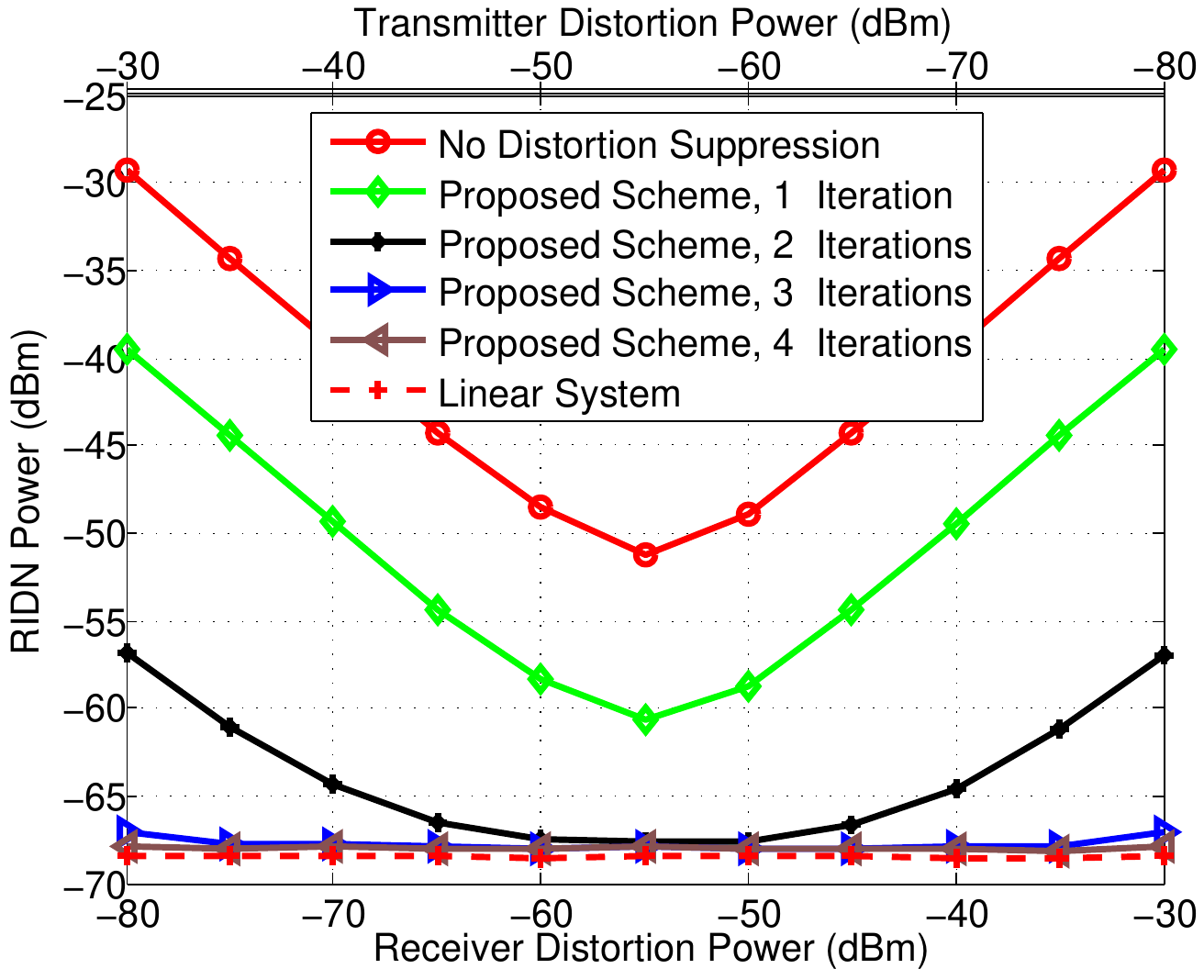}
  \caption{RIDN power at different distortion levels.\label{Fig4Label}}
\end{center}
\end{figure}

In the first simulation scenario, we investigate the performance of the proposed scheme under different transmitter and receiver nonlinearity distortion levels. The target is to evaluate the performance of the proposed scheme under all distortion scenarios: (i) transmitter distortion is greater than receiver distortion, (ii) receiver distortion is greater than transmitter distortion, and (iii) transmitter and receiver distortion are comparable. Figure~\ref{Fig4Label} shows the RIDN power at different transmitter and receiver distortion levels and phase noise power of $-$70dBm. The top and bottom x-axes shows the transmitter and receiver distortion values respectively.

The conclusions from Figure~\ref{Fig4Label} are multifold: first, regardless of the distortion level, the proposed scheme is able to suppress the distortion to the level of the next bottleneck (e.g. phase noise in this case) and achieve a performance that is highly close (less than 0.5dB difference) to the performance of a linear receiver. Second, when the difference between the distortion level and the level of the next bottleneck increases, the number of iterations required to suppress the distortion signal increases. The reason is that each iteration has a limited suppression gain controlled by the channel estimation error, thus more suppression require more iterations. Finally, comparing the left side of Figure~\ref{Fig4Label} to the right side we note that, because the nonlinearity coefficients estimation algorithm adaptively selects the coefficient to start with, the proposed scheme performs the same way whether the transmitter distortion dominates receiver distortion or vise versa.

\begin{figure}[t]
\begin{center}
\noindent
  \includegraphics[width=3.3in,trim= 0in 0in 0in 0in]{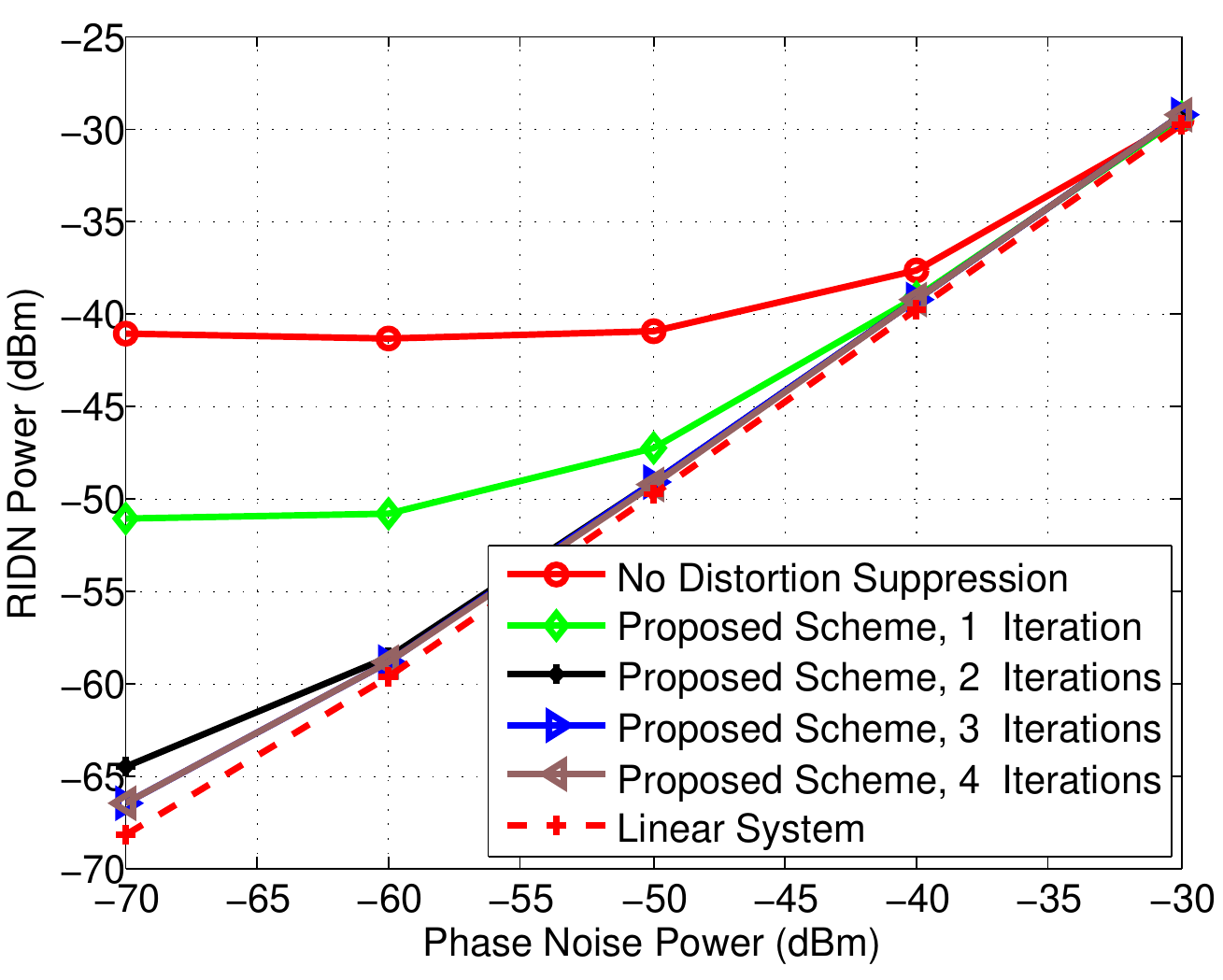}
  \caption{RIDN power at different phase noise levels.\label{Fig5Label}}
\end{center}
\end{figure}

In the previous simulation scenario the system is simulated in the case when the nonlinear distortion dominates other noise components. For complete performance evaluation, the performance is investigated under different phase noise power levels in order to investigate the case when the nonlinear distortion is not the limiting factor. Figure~\ref{Fig5Label} shows the RIDN power at different phase noise levels with a $-$45dBm transmitter and receiver distortion power. The results show that when other noise component dominates nonlinear distortion, the proposed scheme achieves same performance as the case where no distortion suppression is performed. In other words, the proposed scheme does not degrade the performance at low distortion levels.

\begin{figure}[t]
\begin{center}
\noindent
  \includegraphics[width=3.3in,trim= 0in 0in 0in 0in]{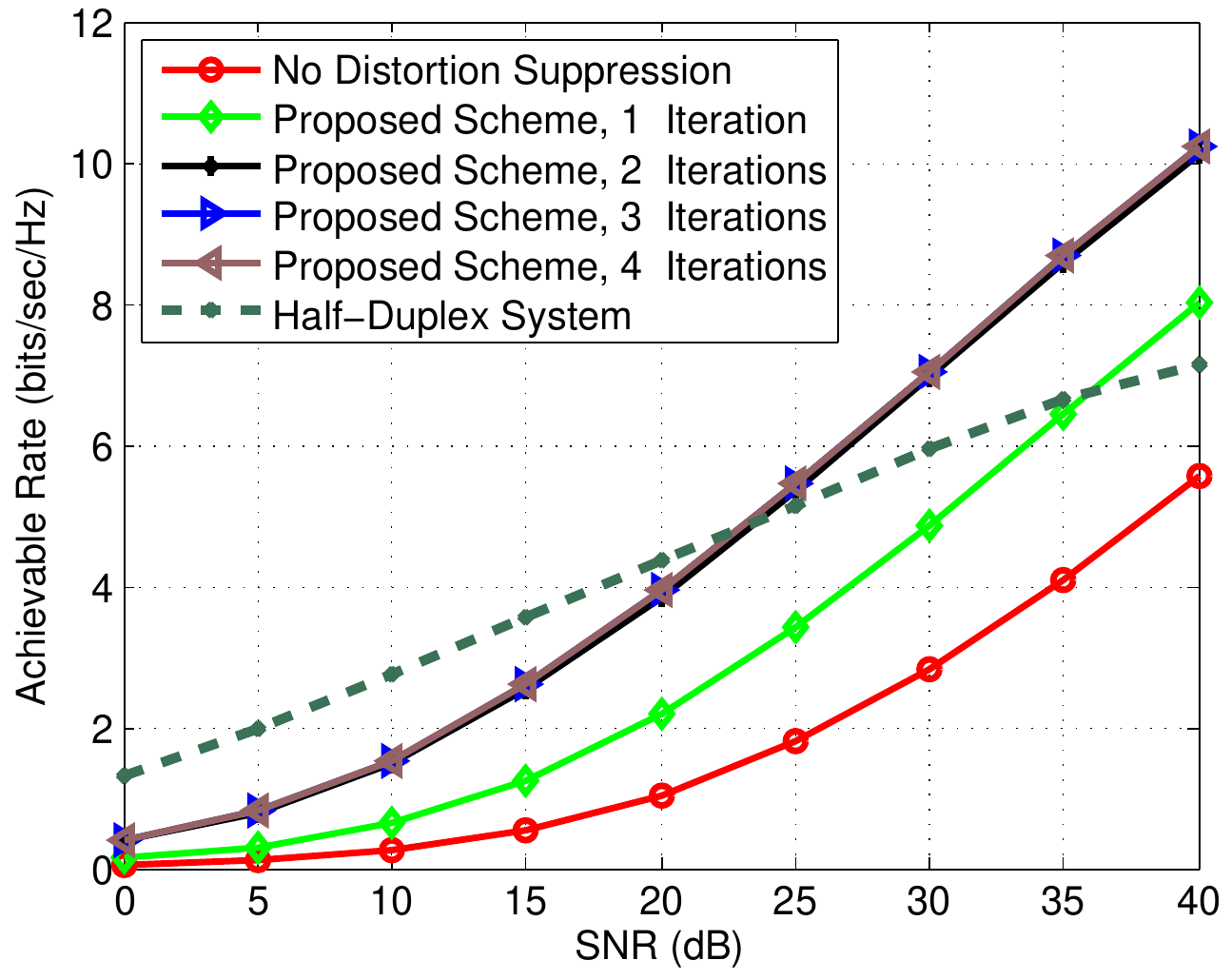}
  \caption{Full-duplex and half-duplex achievable rates at received self-interference signal strength = $-$30dBm, normalized transmitter and receiver distortion power = $-$45dB, and normalized phase noise power = $-$60dB.\label{Fig6Label}}
\end{center}
\end{figure}

In the following simulation scenario, the overall full-duplex system performance is investigated and compared to the corresponding half-duplex system performance. Figure~\ref{Fig6Label} shows the full-duplex and half-duplex system's achievable rate at different half-duplex signal-to-noise ratios (SNR). Since half-duplex system performance is usually limited by the receiver Gaussian noise, the SNR is defined as the received signal-of-interest power divided by the receiver Gaussian noise power. The parameters for this simulation scenario are shown in the figure caption. The results show that when the nonlinear distortion dominates other noise components, performing distortion suppression using the proposed scheme significantly improves the full-duplex system's spectral efficiency and allows full-duplex systems to achieve better rate than half-duplex systems at high SNR scenarios.

\section{Conclusion}
In this paper, a digital-domain self-interference cancellation scheme for full-duplex OFDM systems is proposed. The proposed scheme increases the amount of cancellable self-interference power by suppressing the distortion caused by the transmitter and receiver nonlinearity. The proposed scheme is able to suppress the nonlinear distortion to the level of the next significant noise component, and achieve a performance that is less than 0.5dB off the performance of a linear full-duplex system.

\section{Acknowledgment}
This work is partially supported by Qatar National Research Fund (QNRF) grant through National Priority Research Program (NPRP) No. 4-1119-2-427.


\end{document}